# A Deep Neural Network Model of Particle Thermal Radiation in Packed Bed

HaoWu[2], Shuang Hao[1]
1. School of Computer and Information Technology, Beijing Jiaotong University, Beijing, 100044, China
2. Institute of Nuclear and New Energy Technology, Collaborative Innovation Center of Advanced Nuclear Energy Technology, Key Laboratory of Advanced Reactor Engineering and Safety, Ministry of Education, Tsinghua University, Beijing, 100084, China
wuhao1938@hotmail.com, haoshuang@bjtu.edu.cn

**Abstract**

Prediction of particle radiative heat transfer flux is an important task in the large discrete granular systems, such as pebble bed in power plants and industrial fluidized beds. For particle motion and packing, discrete element method (DEM) now is widely accepted as the excellent Lagrangian approach. For thermal radiation, traditional methods focus on calculating the obstructed view factor directly by numerical algorithms. The major challenge for the simulation is that the method is proven to be time-consuming and not feasible to be applied in the practical cases. In this work, we propose an analytical model to calculate macroscopic effective conductivity from particle packing structures Then, we develop a deep neural network (DNN) model used as a predictor of the complex view factor function. The DNN model is trained by a large dataset and the computational speed is greatly improved with good accuracy. It is feasible to perform real-time simulation with DNN model for radiative heat transfer in large pebble bed. The trained model also can be coupled with DEM and used to analyze efficiently the directional radiative conductivity, anisotropic factor and wall effect of the particle thermal radiation.

## Introduction

Radiative heat transfer widely exists in the large packed beds, including circulating fluidized beds (Borodulya and Kovensky 1983; Hou et al. 2015), high temperature solid particle solar receiver (Johnson et al. 2019) and pebble-bed power plant (Wu et al. 2016). From the aspect of design and engineering applications, it is very meaningful and necessary to simulate the flow and heat transfer processes in packed beds. At early stages, all physical mechanisms involved in a packed bed were not understood thoroughly. The porous model was firstly established to provide an empirical approach to analyze flow and heat transportation at the transient and steady state in pebble bed. The prediction accuracy depends greatly on the empirical parameters.

In Euler-Euler approach, the effect of thermal radiation is considered as the effective thermal conductivity and calculated by empirical correlations. In computational fluid dynamics (CFD) - discrete element method (DEM) framework, which investigate two-phase flow behaviors in particle scale level, particle motion, particle-fluid, contact conduction and convection without thermal radiation are discussed in numerical simulations (Kloss et al. 2012; Patil et al. 2015). And the radiative heat transfer also is not included in the analysis of entransy dissipation (Wang et al. 2019).

From results of the heat transfer experiments (Abou-Sena et al. 2005; Van Antwerpen et al. 2010), radiative thermal conductivity is an important part and increases significantly with the operation temperature. Moreover, the radiation exchange factor is well-known as its non-dimensional parameter, and widely used for analysis of particle radiation. There is still lacking a physical expression of the radiation exchange factor based on the particle packing structure. In DEM, all particles motion is handled by the contact force to others and physical walls. Similarly, the conductive heat transfer is calculated by the contact thermal resistance. However, the computation of thermal radiation in packed bed becomes time-consuming. The numerical model often applies a short-range energy cutoff (Cheng and Yu 2013). In (Johnson et al. 2019), the obstructed view factor is simplified to a function of the distance. Obviously, the local packing characteristics is not considered. In radiation transfer equation (RTE), the particle system is assumed isotropic homogeneous medium with scattering, emitting and absorption. It is the scattering coefficient and absorption coefficient is applied in the numerical model. Thus, it is difficult to discuss the effect of the physical properties such as surface emissivity and solid conductivity.

The obstructed view factor between spheres in packed bed is a bridge between particle-scale packing and macroscopic properties including radiative thermal conductivity, anisotropic factor and wall effect, even for the radiation between complex non-spherical particles Traditionally, the view factor is calculated by numerical integration or Monte Carlo method (Walker, Xue and Barton 2010). However, the computational time is unacceptable to obtain satisfactory result in large packed beds. Thus, to analyze and discuss the long-range radiation in packed bed, the main task is developing an efficient approach to calculate the obstructed view factor.

Artificial neural network (ANN) with multilayer perceptron provides a powerful predictor for regression and classification. The network with deep structure can used to appropriate the non-linear function of the view factor. It can be trained by efficient backpropagation algorithm (Foresee and Hagan 1997; Wang, Sun and Xu 2019) with big data with a huge number of view factor case.

**Contribution**. Our contributions are Summarized below.
➢ We describe a numerical model of radiative heat transfer in packed bed and analytical expressions between packing and macroscopic properties.
➢ We generate a large dataset with different view factor cases for the function regression.
➢ We develop a deep neural network (DNN) model with reasonable structures to calculate the view factor in packed bed efficiently and accurately.
➢ With trained DNN model, we perform real-time numerical simulation of radiative heat transfer in packed bed.

## Related Work

The artificial neural network (ANN) now is regards as a versatile tool of machine learning in many engineering cases. For fluid flow and heat transfer, a fully connected neural network model is trained in (Alemany et al. 2019) to predict the trajectory of hurricanes. Data-based approach of ANN is reported by (Beck, Flad and Munz 2019) to model the turbulence in large eddy simulation. (Chang et al. 2018) presents ANN model to predict heat transfer behaviors of the supercritical water. The model is trained by 5280 data points, which is collected from published experimental measurements. The results show that the ANN performance is considerably better than the empirical correlations. ANN is also applied to predict the physical properties of alkanes (Santak and Conduit 2019), convective heat transfer of supercritical carbon dioxide (Ye et al. 2019) and pool boiling (Zendehboudi and Tatar 2017).

In discrete particle simulation, (Benvenuti, Kloss and Pirker 2016) develop artificial neural networks to link macroscopic experiments to simulation parameters. (Kumar et al. 2018) employees ANN to predict mass discharge rate from conical hoppers. The input parameters include bulk density, internal angle, particle diameter, friction coefficient. And the data for training is generated by DEM simulation. In (Desu, Peeketi and Annabattula 2019), ANN model with 3 hidden layers is trained by 11-dimensional data from resistor network (RN) model to predict effective thermal conductivity of thermal conduction. The presence stagnant gas and the Smoluchowski effect are discussed for the conductive heat transfer and the model is much faster than traditional method with good accuracy. But no thermal radiation term is considered in discussion.

The heat transfer experiments of packed beds under high temperature ranges are reported in (Wakao and Kato 1969), (Nasr, Viskanta and Ramadhyani 1994) and (Earnshaw, Londry and Gierszewski 1998) and aimed to measure the total effective thermal conductivity, which mainly includes the radiative part and the conductive part. The materials for the particles of the bed include glass, aluminum and lithium zirconate. The experimental results agree generally with the Zehner–Schlunder correlation and Kunii–Smith correlation.

Physically, the conductivity of the thermal radiation increases with the average particle size greatly (Fillion et al. 2011). Commonly, the particle size in many cases of the packed beds is in range of 1.0 ~ 10.0 mm and it is much smaller than the pebbles of 60 mm in diameter in pebble bed. In recent years, the measurements of pebble beds operated under the similar conditions of the nuclear reactor are conducted in the high temperature test unit (HTTU) (Rousseau et al. 2014) and test facility for pebble bed equivalent conductivity measurement (TF-PBEC) (Ren et al. 2017). The particles are spheres of graphite. From the empirical correlations reviewed by (Tsotsas and Martin 1987), the effect of emissivity on the radiation exchange factor is a separable term in the radiative flux equation and it is a continuous monotonically increasing function in the dense bed of spheres. When the emissivity is zero, only scattering happens between spheres and the net radiative flux becomes zero.

## Methodology

**Directional radiative conductivity**
For the two-phase flow and heat transfer in packed bed, the CFD equation is written as

$$\frac{\partial(\rho_f \alpha_f)}{\partial t} + \nabla \cdot (\rho_f \alpha_f \mathbf{u}_f) = 0 \quad (1)$$

$$\frac{\partial(\rho_f \alpha_f \mathbf{u}_f)}{\partial t} + \nabla \cdot (\rho_f \alpha_f \mathbf{u}_f \mathbf{u}_f) = -\alpha_f \nabla p + \nabla \cdot (\mu_f \alpha_f \nabla \mathbf{u}_f) + S_m \quad (2)$$

$$\frac{\partial(\rho_f C_p \alpha_f T_f)}{\partial t} + \nabla \cdot (\rho_f C_p \alpha_f \mathbf{u}_f T_f) = \nabla \cdot (\lambda_f \alpha_f \nabla T_f) + S_e \quad (3)$$

where $\alpha_f$ and $\rho_f$ are the local porosity and fluid density. $\mathbf{u}_f$ and $T_f$ are the fluid velocity and temperature. $\mu_f$ and $p$ are the fluid viscosity and the pressure. $\lambda_f$ and $C_p$ are the thermal conductivity and specific heat. $S_m$ and $S_e$ are the source terms for particle-fluid interaction of drag force and forced convection.

For particle motion, packing and heat transfer are usually calculated by the discrete element method (DEM) (Zhu et al. 2007). The basic governing equations are

$$m_i \frac{d^2 \mathbf{X}_i}{dt^2} = \sum_{k=1}^n \mathbf{F}_{t,ik} + \sum_{k=1}^n \mathbf{F}_{n,ik} + m_i \mathbf{g} \quad (4)$$

$$I_i \frac{d\boldsymbol{\omega}_i}{dt} = \mathbf{R}_i \times \sum_{k=1}^n \mathbf{F}_{t,ik} + \mathbf{M}_r \quad (5)$$

$$m_i C_{p,i} \frac{dT_i}{dt} = Q_{cond} + Q_{conv} + Q_{rad} \quad (6)$$

where $\mathbf{X}_i$ and $T_i$ are the particle velocity and temperature at time $t$. $m_i$, $I_i$ and $C_{p,i}$ are the particle mass, momentum of inertia and specific heat respectively. $\boldsymbol{\omega}_i$ and $\mathbf{M}_r$ are angular velocity and the rolling friction torque. $\mathbf{F}_{t,ik}$ and $\mathbf{F}_{n,ik}$ are the tangential contact force and normal contact force. $\mathbf{g}$ is the gravity vector. $Q_{cond}$, $Q_{conv}$ and $Q_{rad}$ are the conduction flux, heat convection and thermal radiation flux respectively. At high temperature, such as the 1000°C ~ 1200°C in experiments, radiative heat transfer becomes a dominant part. The radiation flux is determined by the temperature difference between the particle and all its surrounding ones. The gray-body radiative flux from particle $i$ to particle $k$, which are positioned in $P_i = (x_i, y_i, z_i)$ and $P_k = (x_k, y_k, z_k)$, is formulated as

$$\vec{q}_{ik} = \varepsilon_r \sigma A_i V_{ik}(T_i^4 - T_k^4) \cdot \vec{n}_{ik} \quad (7)$$

where $A_i$, $\varepsilon_r$ and $\sigma$ are particle surface area, emissivity and Stefan–Boltzmann constant. $V_{ik}$ is the obstructed view factor to surroundings and $Q_{rad} = \sum_{k=1}^N |\vec{q}_{ik}|$. The direction vector is defined as $\vec{S}_{ik} = P_k - P_i$ and unit vector is given as $\vec{n}_{ik} = \vec{S}_{ik}/|\vec{S}_{ik}|$. For the bed of anisotropic packing, the case heat transfer in a 1D thick plate is applied to obtain the effective thermal conductivity (ETC). If the temperature gradient is in the direction $\vec{\beta} = (\beta_x, \beta_y, \beta_z)$, which is a unit vector i.e. $|\vec{\beta}| = 1$, and particle $i$ is in its central line, the steady temperature distribution with uniform heat source is written as

$$T(\vec{u}_{ik}) = T_i - \frac{q_s}{2k_\beta} \vec{u}_{ik}^2 \quad (8)$$

where $q_s$ is the heat transfer power. $\vec{u}_{ik}$ is the projection vector of $\vec{S}_{ik}$ in direction $\vec{\beta}$ (see Figure 1) and it is $\vec{u}_{ik} = (\vec{S}_{ik} \cdot \vec{\beta}) \cdot \vec{\beta}$. $k_\beta$ is the directional radiative conductivity and it being the ETC in direction $\vec{\beta}$.

From Eq. (7) and Eq. (8), in which it is $T_i^4 - T_k^4 = 4T^3(T_i - T_k)$ at low temperature gradients, $k_\beta$ for bed of mono-sized spheres is derived as

$$k_\beta = 12\sigma\varepsilon_r T^3 \frac{(1-\varphi)}{Nd} \sum_{k=1}^N \sum_{i=1}^N V_{ik} |\vec{S}_{ik}|^2 (\vec{n}_{ik} \cdot \vec{\beta})^2 \quad (9)$$

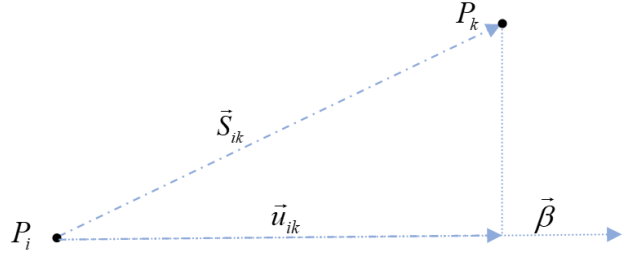

Figure 1 The position of particle $i$, $k$ and direction $\vec{\beta}$.

where $\varphi$, $N$ and $d$ are the average porosity, particle number and the diameter in the packed bed. It can be found the effective thermal conductivity is directly proportional to the cube of particle absolute temperature. And its dimensionless parameter is

$$F_\beta = \frac{k_\beta}{4\sigma d T^3} = 3\varepsilon_r \frac{(1-\varphi)}{N} \sum_{k=1}^N \sum_{i=1}^N V_{ik} |\vec{h}_{ik}|^2 (\vec{n}_{ik} \cdot \vec{\beta})^2 \quad (10)$$

where $F_\beta$ is the directional radiation exchange factor and $\vec{h}_{ik} = \vec{S}_{ik}/d$.

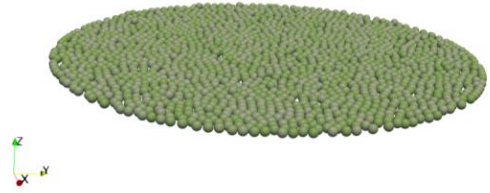

Figure 2 A thin packed bed only with height of one-layer sphere.

For one-layer bed shown in Figure 2, it is $\vec{n}_{ik} \cdot \vec{\beta} = 0$ or all sphere pair and the directional radiative conductivity in $z$ axis is zero. However, it is greater than 0 in other directions, i.e. $F_x > 0$, $F_y > 0$ and $F_z = 0$. Moreover, it can be found that the packing is isotropic approximately in $x$-$y$ plane. To describe the directional radiative conductivity along a plane $\gamma$, the case in a cylindrical coordinate is applied and particle $i$ still is in its central line. In this case, the temperature with uniform heat source is

$$T(\vec{\rho}_{ik}) = T_i - \frac{q_s}{4k_\rho} \vec{\rho}_{ik}^2 \quad (11)$$

where $k_\rho$ is the directional radiative conductivity and $\vec{\rho}_{ik}$ is the projection vector in the plane and it is shown in Figure 3. The vector is $\vec{\rho}_{ik} = \vec{S}_{ik} - (\vec{S}_{ik} \cdot \vec{n}_p) \cdot \vec{n}_p$ and $\vec{n}_p$ unit normal vector of the plane. $k_\rho$ is derived as

$$k_\rho = 6\sigma\varepsilon_r T^3 \frac{(1-\varphi)}{Nd} \sum_{k=1}^N \sum_{i=1}^N V_{ik} \vec{\rho}_{ik}^2 \quad (12)$$

Finally, if the packing of the packed bed is isotropic in all directions, the average effective thermal conductivity is written as

$$k_v = 4\sigma\varepsilon_r T^3 \frac{(1-\varphi)}{Nd} \sum_{k=1}^N \sum_{i=1}^N V_{ik} \vec{S}_{ik}^2 \quad (13)$$

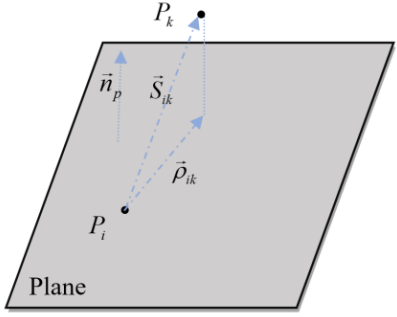

Figure 3 The positions of plane and particles.

The anisotropic factor, which is a dimensionless parameter for quantifying the degree of anisotropic directional radiative conductivity in packed bed, is defined as

$$\eta = \frac{k_{\beta,\max}-k_{\beta,\min}}{k_{\beta,\max}} = \frac{F_{\beta,\max}-F_{\beta,\min}}{F_{\beta,\max}} \quad (14)$$

where $F_{\beta,\max}$ and $F_{\beta,\min}$ are the maximum and minimum in all directions. For the case in Figure 2, it becomes $F_{\beta,\max} = F_z = 0$ and $\eta = 1$. For isotropic case, it will be $F_{\beta,\max} = F_{\beta,\min}$ and $\eta = 0$. When $\eta$ decreases from 1 to 0, the degree of anisotropy decreases to be isotropic.

In the uniform continuum model (Wu et al. 2019), the radiative heat transfer is formulated as

$$\rho_0 \varepsilon_r A_i \sigma \left[ \int_{\mathbf{R}^3} K(|\mathbf{x}-\mathbf{x}'|) T^4(\mathbf{x}') d\mathbf{x} - T^4(\mathbf{x}) \right] + q = 0 \quad (15)$$

where $\rho_0$ is the number density and $q$ is the heat source term. $K(\mathbf{x})$ is the kernel function and directly determined by the radial distribution function and the view factor.

**View factor function**

For a practical case shown in Figure 4, in which simulations are performed by DEM, the spheres are of 60 mm in diameter and packing density is 0.61. The calculation of view factor $V_{ik}$ between all sphere pairs costs most computational time in packed bed. The view factor between two spheres is expressed as

$$V_{ik} = \frac{1}{\pi A_i} \int_{A_k} \int_{A_i} \frac{\cos\alpha_i \cos\alpha_k}{r^2} \upsilon(\mathbf{x}_i, \mathbf{x}_k) dA_i dA_k \quad (15)$$

where $\upsilon(\mathbf{x}_i, \mathbf{x}_k)$ is the visibility function. For $\upsilon(\mathbf{x}_i, \mathbf{x}_k) = 1$, the ray from particle $i$ can reach the particle $j$. In other cases, $\upsilon(\mathbf{x}_i, \mathbf{x}_k)$ is 0. The view factor from the sphere to physical walls can be calculated from the result to other spheres, and it is can be written as

$$V_{i,w} = 1 - \sum_{k=1}^{N} V_{ik} \quad (17)$$

For two spheres without obstruction by other ones, the analytical expression is written as

$$V_{ik} = \frac{1}{\pi |\vec{h}_{ik}|} \int_0^{\pi/2} \frac{2\zeta}{\sqrt{\vec{h}_{ik}^2 - 4\cos\zeta}} \sin 2\zeta \, d\zeta \quad (18)$$

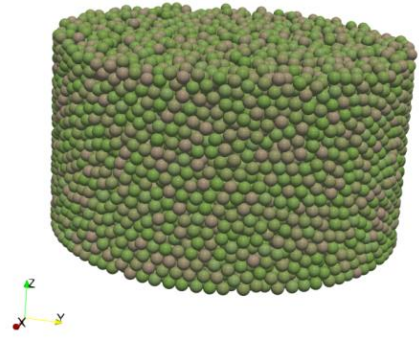

Figure 4 The numerical result of particle packing in a cylindrical packed bed.

For the cases in Figure 4, the sphere pair may be obstructed by other spheres. The view factor function becomes much more communicated and non-linear. It is necessary to use numerical procedures general complex cases. Traditionally, the view factor in packed bed is obtained by Monte Carlo method (MCM) by using a ray tracing approach. The advantages of MCM are of low memory cost and efficient for multithreading. The disadvantage is that many rays need to be traced for achieving good accuracy. From numerical tests, it is good enough to apply $1\times10^8$ rays to calculate view factor in packed bed. The average particle number satisfying $V_{ik} > 0$ is about 130 in the simulation. The average time of the calculation for a sphere is about 38.8s in serial mode, in which the simulation is performed in the Intel Core i7-8700K. The time decreases to 7.26s per particle with 6 processors. With GPU accelerating, the simulations are performed in Nvidia GeForce GTX 1070, the average time to calculate view factor to surroundings decreases greatly to 0.64s per particle. However, neither CPU nor GPU performance is capable for the practical cases. As the fact that there are $2.7\times10^4$ spheres in the experimental nuclear reactor and $4.2\times10^5$ spheres in the demonstration power plant. For a packed bed of $1\times10^5$ spheres, it takes about 201.7h with CPU and 17.8h with GPU.

**Deep neural network**

To accelerate the calculation of the view factor, a deep neural network model, which is a universal estimator for function regression, is applied to understand the rule to calculate view factor between all spheres in packed bed. The structure of the model is shown in Figure 5.

A large dataset calculated by the GPU calculation is used as the input for training. Then the preprocessing of the data is to delete unrelated points, which are far away from the particle pair and make no contribution to the view factor. The principal component analysis (PCA) is used for the dimensionality reduction. The neural network with three hidden layers is applied for the regression and every hidden layer contains up to 30 neurons. The connections are from the input and every previous layer to following layers. The data is divided randomly into 80% for training, 10% for validation and 10% for testing. The model is trained by the algorithm of Bayesian regularization back-propagation (Foresee and Hagan 1997).

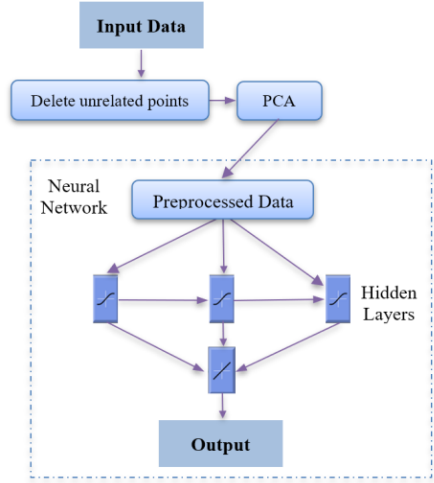

Figure 5 The structure of the neural network model for calculating view factor.

## Experiments

### Experimental setup

**Datasets.** No large dataset of view factor in packed bed is reported in related literatures. We generate the view factor dataset by following steps: (1) Divide the view factor function into 11 groups. For case with two spheres, it is $V_{12} = f_2 = f(P_1, P_2)$. For 3 spheres case, the function is $V_{12} = f_3 = f(P_1, P_2, P_3)$. And the function $V_{12} = f_{12} = f(P_1, P_2, ... P_{12})$ is the case of 12 spheres, which means that the ray from $P_1$ to $P_2$ may be affected 10 spheres ($P_3$, $P_4$, ..., $P_{12}$). For the cases of more than 10 spheres closest to the line $P_1P_2$, it is reasonable to consider only top 10 ones. (2) For the function $f_2 = f(P_1, P_2)$, the view factor is calculated directly by Eq. (18). For $f_3, f_4, ..., f_{12}$, generate random points $P_1, P_2, P_3 ...$ of mono-sized spheres without overlapping. Calculate the view factor by MCM with GPU accelerating. (3) All the cases are in the range of $|P_1 - P_2| < 6d$. The dataset is stored in the form $P_1, P_2, ..., V_{12}$ and the record number is listed in Table 1.

Table 1 The dataset for training the view factor function.

| Function | Record number | Function | Record number |
|---|---|---|---|
| $f_3$ | 4.0×10⁴ | $f_8$ | 1.4×10⁵ |
| $f_4$ | 5.5×10⁴ | $f_9$ | 1.4×10⁵ |
| $f_5$ | 8.9×10⁴ | $f_{10}$ | 1.4×10⁵ |
| $f_6$ | 1.4×10⁵ | $f_{11}$ | 1.6×10⁵ |
| $f_7$ | 1.4×10⁵ | $f_{12}$ | 1.6×10⁵ |

**Training.** The preprocessing of the data can be performed by translation, scale and rotation. For the points $P_1, P_2, P_3 ...$, we put $P_1$ at origin and $P_2$ at $x$ axis. $P_3$ is in $x$-$y$ plane and all points are scaled by particle radius. Then every column for training is normalized in [-1, 1] We use mean square error (MSE) as loss function and it is given as

$$\text{MSE} = \frac{1}{N_r}\sum_i^{N_r}(\hat{V}_i - V_{12})^2 \quad (19)$$

where $N$ is the record number of the dataset for training. $V_{12}$ is the input value and $\hat{V}_i$ is the predicted view factor. The maximum of view factor is about 0.07558 for the case without obstruction and all dataset is clean. Thus, the iteration will stop at the absolute error $e = |\hat{V}_i - V_{12}| < 5 \times 10^{-5}$ for 90% dataset.

**Baselines.** For the function of two spheres $f_2 = f(P_1, P_2)$, a feedforward neural network with a single hidden layer of only 10 neurons is applied and trained by 315 data points. The prediction error is $8 \times 10^{-6}$ and the performance is much better than that of linear regression and symbolic regression. The model can be used in the application to avoid the calculation of the complex the integration term in Eq. (18). Moreover, for the view factor function $f_6 = f(P_1, P_2, P_3, P_4, P_5, P_6)$, the results with difference structures are given in Table 1. It is found that the error decreases with increasing of the layer number and neuron number. The performance is also improved with preprocessing. When the network structure increases to 25×25×25, the error with preprocessing decreases to acceptable value of 2.7×10⁻⁵. And for $f_{12}$, the neurons in network increases to 40 for good prediction.

Table 2 The structure of neural network for the function $f_6$

| Model | Neurons | Preprocessing | Average error |
|---|---|---|---|
| A | 10 | Yes | 1.5×10⁻³ |
| B | 10×5 | Yes | 1.2×10⁻³ |
| C | 10×10 | No | 9.6×10⁻⁴ |
| D | 10×10 | Yes | 3.3×10⁻⁴ |
| E | 25×25×25 | Yes | **2.7×10⁻⁵** |

### Experimental results

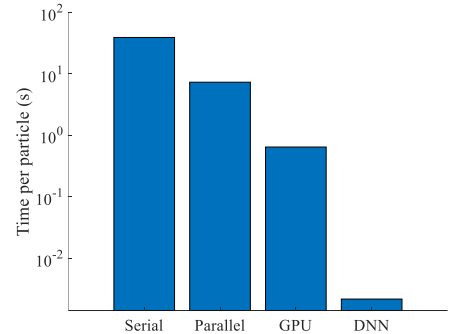

Figure 6 The comparison of calculation speed for view factor with different method.

**Prediction performance.** The application of the trained deep neural network model in packed bed is conducted by following procedures. First, find all surrounding spheres of particle $i$. The neighbors are searched by $K$d-tree in the range of 6 times of the particle radius. The view factor decreases to 0 approximately for the pairs over 6 times of the particle radius and it can be neglected in the discussion; Then, Apply the trained deep neural network (DNN) model of the function regression to calculate the view factor; Finally, assemble the all view factor as a sparse matrix.

By testing the same case, average computational time shown in Figure 6 is $2.15\times10^{-3}$s per particle using DNN. It is 298 times faster than that of GPU, in which DNN prediction also achieves good accuracy shown in Figure 7, and only takes about 3.6 min to calculate view factor between spheres in packed bed of $1\times10^5$ spheres, which makes feasible to perform real-time simulation of radiative heat transfer.

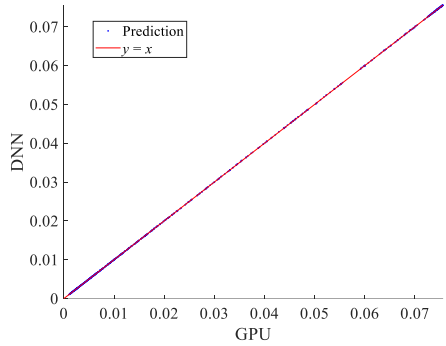

Figure 7 Prediction accuracy of the trained deep neural network for the view factor.

**View factor enclosure.** For a sphere in packed bed, accumulated view factor is sum of all view factor to surroundings in range of a given distance. The average statistical results with DNN model are shown in Figure 8.

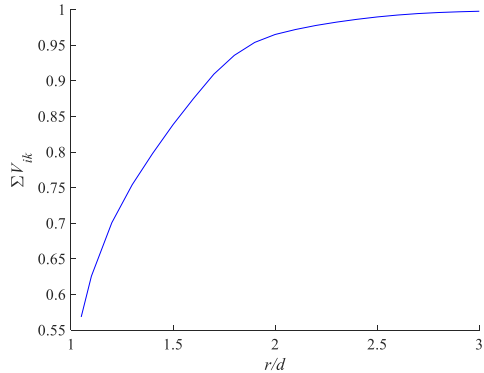

Figure 8 Average accumulated view factor with distance for spheres in packed bed.

The value increases with the distance and strictly ranged of 0 to 1. In $1.05d$ that is very near to the particle, $\sum V_{ik}$ is about 0.57. It is because there are about 7~8 spheres contact directly with view factor of 0.07558. $\sum V_{ik}$ increases greatly with the distance at $r \leq 2d$ and slowly reaches its maximum at about $3d$. In $1.5d$, which is discussed in short range model, accumulated view factor reaches to 0.84. Thus, it is not accurate enough for the model only considering the pairs in the distance of 1.5 times particle diameter. In the full range of $3.0d$, $\sum V_{ik}$ is 0.9977 and it is reasonable to be modeled as an enclosed space in the numerical simulation.

**Anisotropic factor.** For directional radiative conductivity in $x$, $y$ and $z$ axis, $F_\beta$ becomes $F_x$, $F_y$ and $F_z$, in which direction $\vec{\beta}$ is $(1,0,0)$, $(0,1,0)$ and $(0,0,1)$ respectively. The position vector can be written as $\vec{h}_{ik} = a_{ij}\vec{n}_x + b_{ij}\vec{n}_y + c_{ij}\vec{n}_z$. It can be proven that the relationships between radiation exchange factor in a given direction and its average in a plane or the whole bed are $F_{xy} = (F_x + F_y)/2$ and $F_v = (F_x + F_y + F_z)/3$. For cylindrical bed, the radiation exchange factor almost keeps a constant under different polar angles. From results under different azimuthal angle $\varphi$, $F(\varphi)$ can be formulated as $a + b\sin^2\varphi$. The parameter $b$ decreases with the height of the bed and it will be isotropic at $b = 0$.

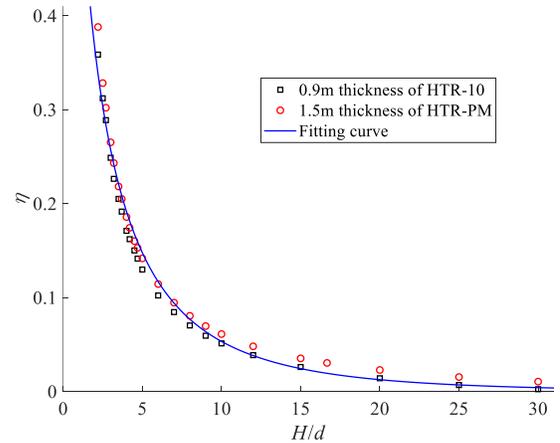

Figure 9 The anisotropic factor under different size of the pebble bed.

The results of anisotropic factor are given in Figure 9 for HTR-10 of 0.9m in thickness and HTR-PM of 1.5m in thickness. Commonly, the fitting curve is given as

$$\eta = 1.73\exp(-1.1\sqrt{H/d}) \quad (20)$$

where $H$ is the packing height. At height of $6.5d$, the anisotropic factor decreases to about 0.1. At $H = 1.0$m, $\eta$ are 0.022 for HTR-10 and 0.031 for HTR-PM.

**Wall effect.** In uniform continuum model, the view factor from sphere to wall can be expressed as

$$X_w = 1 - \int_{-\infty}^{x_w}\int_{-\infty}^{+\infty}\int_{-\infty}^{+\infty} K(\mathbf{x}) dx dy dz \quad (21)$$

where $x_w$ is the distance to the wall. In discrete model, the view factor can be obtained from sphere-sphere data in Eq (17).

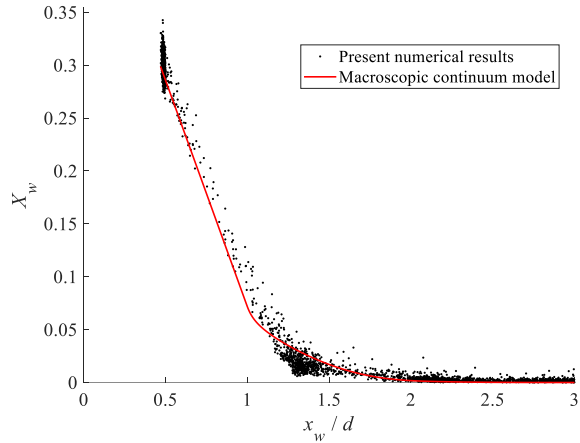

Figure 10 The view factor from the sphere to physical wall in pebble bed

The numerical results of the wall effect are given in Figure 10. Generally, the wall effect decreases with the distance and the continuum model is the average value. For $x_w < d$, it is the wall region and the view factor to the wall is greater than that to spheres. At $d \leq x_w < 2.5d$, it is near-wall region and the wall effect is much less than the wall region. At $x_w \geq 2.5d$, it is the bulk region and wall effect can be neglected.

## Conclusions

For the thermal radiation in packed bed under high temperature, a deep neural network model is applied to investigate the directional radiative conductivity, anisotropy and wall effect. It is found that:

(1). The radiative heat transfer in pack bed is of long-range. Directional radiative conductivity is a macroscopic parameter in a given direction. The calculation is related with the obstruction view factor.

(2). The deep neural network (DNN), which is trained by big data, is good predictor of the view factor. The calculation speed is significantly accelerated, and it makes feasible to perform real-time simulation of the radiative heat transfer.

(3). From numerical results, the anisotropic factor decreases with the packing height. With the ANN model and uniform continuum model, the wall effect on the thermal radiation can be 3 parts of wall region, near-wall region and bulk region.